\documentstyle[prb,twocolumn,aps,psfig,amssymb,floats]{revtex}

\newcommand{\partabl}[2]{\frac{{\rm \partial} #1}{{\rm \partial} #2}}

\begin{document}
\draft

\twocolumn[\hsize\textwidth\columnwidth\hsize\csname @twocolumnfalse\endcsname

%%%%%%%%%%%%%%%%%%%%%%%%%%%%%%%%%%%%%%%%%%%%%%%%%%%%%%%%%%%%%%%%%%%%%%%
%%%%%%   TITLE   %%%%%%%%%%%%%%%%%%%%%%%%%%%%%%%%%%%%%%%%%%%%%%%%%%%%%%
%%%%%%%%%%%%%%%%%%%%%%%%%%%%%%%%%%%%%%%%%%%%%%%%%%%%%%%%%%%%%%%%%%%%%%%

\title{Semiclassical approach to calculating the influence of local
lattice fluctuations on electronic properties of metals}
\author{Stefan Blawid} \address{Institut f\"ur Mathematische Physik,
TU Braunschweig, Mendelssohnstr. 3, 38106 Braunschweig}
\author{Andreas Deppeler}
\address{Center for Materials Theory, Department of Physics \&
Astronomy, Rutgers University, 136 Frelinghuysen Road, Piscataway, NJ
08854} \author{A.J.~Millis} \address{Department of Physics, Columbia
University, 538 W. 120th St., New York, NY 10027}

\date{\today}

\maketitle

%%%%%%%%%%%%%%%%%%%%%%%%%%%%%%%%%%%%%%%%%%%%%%%%%%%%%%%%%%%%%%%%%%%%%%%
%%%%%%   ABSTRACT   %%%%%%%%%%%%%%%%%%%%%%%%%%%%%%%%%%%%%%%%%%%%%%%%%%%
%%%%%%%%%%%%%%%%%%%%%%%%%%%%%%%%%%%%%%%%%%%%%%%%%%%%%%%%%%%%%%%%%%%%%%%

\begin{abstract}
We propose a new semiclassical approach based on the dynamical mean
field theory to treat the interactions of electrons with local lattice
fluctuations. In this approach the classical (static) phonon modes are
treated exactly whereas the quantum (dynamical) modes are expanded to
second order and give rise to an effective semiclassical potential. We
determine the limits of validity of the approximation, and demonstrate
its usefulness by calculating the temperature dependent resistivity in
the Fermi liquid to polaron crossover regime (leading to `saturation
behavior') and also isotope effects on electronic properties including
the spectral function, resistivity, and optical conductivity, problems
beyond the scope of conventional diagrammatic perturbation theories.
\end{abstract}


%71.10.-w Theories and models of many electron systems
%71.10.Fd Lattice fermion models (Hubbard model, etc.)
% 71.10.Hf Non-Fermi-liquid ground states, electron 
%          phase diagrams and phase transitions in model 
%          systems
%71.20.Eh Rare earth metals and alloys
%71.27.+a Strongly correlated electron systems; heavy fermions
%71.30.+h Metal-insulator transitions and other electronic transitions
%72.15.-v Electronic conduction in metals and alloys
%71.45.Lr Charge-density-wave systems (see also 75.30.F 
%          Spin-density waves)
%72.80.-r Conductivity of specific materials (for conductivity of
%         metals and alloys, see 72.15)

\vskip2pc]

\section{Introduction}
\label{sec:intro}

The present-day understanding of electron-lattice interactions in
metals is based on the assumption that typical phonon frequencies
$\Omega$ are small relative to typical electron energies $t$, so that
electronic properties may be calculated as an expansion in the
adiabatic parameter $\gamma = \Omega/t$. This was first exploited by
Migdal and Eliashberg (ME)\cite{migdal58} to derive a self-consistent
set of equations for electron and phonon self-energies. Since ME
theory neglects order $\gamma$ vertex corrections it wrongly predicts
a zero result for most isotope effects on electronic properties
(notable exceptions being the superconducting \cite{mcmillan68} and
charge-ordering\cite{blawid01} transition temperatures). Furthermore,
ME theory assumes that the underlying electronic groundstate can be
described by Fermi liquid theory.  However, if the electron-lattice
coupling strength $\lambda = \Lambda/t$ exceeds a critical value (of
order 1) the conduction electrons are believed to form ``small
polarons'',\cite{alexandrov94} so that the electronic groundstate is
fundamentally reconstructed. In this case, although an expansion in
$\gamma$ exists, the starting point is not clear.\\

Signatures of polaronic behavior have been observed in certain complex
narrow band materials. Well-known examples are phonon mediated
superconductors with high transition temperatures, like $\rm BiO$
based superconductors,\cite{puchkov96} alkali-doped $\rm C_{60}$
fullerides,\cite{degiorgi94} and the A15 compounds,\cite{fisk76} and
`colossal magnetoresistance' (CMR) manganites.\cite{okimoto00} Despite
much work many questions remain open: as examples we cite a possible
`saturation' of the resistivity at high temperatures,\cite{fisk76}
unexpected midinfrared transitions in the optical
conductivity,\cite{puchkov96,degiorgi94,okimoto00} and large isotope
effects on electronic properties. For instance, the CMR manganite $\rm
(La_{0.25}Pr_{0.75})_{0.7}Ca_{0.3}MnO_3$ undergoes a metal-insulator
transition upon $\rm ^{16}O$ - $\rm ^{18}O$ isotope
substitution.\cite{belova00} A calculational technique which can
address with these issues is urgently needed.\\

The interaction of conduction electrons with {\it local} lattice
fluctuations may be described by the dynamical mean field theory
(DMFT).\cite{georges96} In this approximation the electron self energy
is momentum independent. Since the full frequency dependence is
retained the corresponding one-electron Green function shows both
coherent and incoherent features. Recent applications of DMFT to the
electron-phonon problem include a systematic expansion in powers of
$\gamma$,\cite{deppeler00} several weak-coupling expansion
schemes,\cite{blawid01,freericks,ciuchi99} and a ``classical''
($\gamma = 0$) study of the Fermi liquid to polaron crossover.
(Treating the $\gamma = 0$ limit without assuming DMFT has not been
possible yet.\cite{kabanov93}) The expansion of
Ref. \onlinecite{deppeler00} is restricted to very low temperatures
and cannot access the intermediate or strong coupling regimes where
important physics such as resistivity saturation or polaronic effects
is expected. Because of the mismatch between phonon and electron
energy scales numerical simulations\cite{freericks} are difficult to
implement accurately, except in the ``antiadiabatic'' limit $\gamma
\sim 1$ of unclear physical relevance.\\

Here we propose a new ``semiclassical'' method for treating
intermediate and strong couplings in the physically relevant limit
$\gamma \ll 1$.  We treat the classical (static) phonon modes exactly
(as in Ref.\onlinecite{millis96}) and expand the quantum (dynamical)
modes to second order. The resulting action consists of a classical
part plus an effective semiclassical potential. The method is valid
for a wide range of electron-phonon couplings and resulting
groundstates. Here we consider only electron-phonon interactions
(i.e. the Holstein model). However, the method can be generalized
(along the lines of Ref.\onlinecite{deppeler00,deppeler02}) to study
phonon effects in the presence of strong electron-electron
interactions. We focus on the crossover regime between Fermi liquid
and polaron physics which was not easily tractable by previous
methods. The remainder of the paper is organized as follows. In
Sec. II we introduce the Holstein model and the semiclassical
method. In Sec. III we discuss additional simplifications which are
convenient for numerically implementing the method. Sec. IV studies
analytical limits and defines the observables (spectral function,
resistivity and optical conductivity) we consider. Numerical results
are presented and discussed in Sec. V. We conclude in Sec. VI.\\

\section{Model and Formalism}

In this paper we study the Holstein model\cite{holstein59} of
electrons interacting with local lattice fluctuations: $H_{\rm hol} =
H_{\rm el} + H_{\rm ph} + H_{\rm el-ph}$ with
\begin{eqnarray}
H_{\rm el} & = & -\sum_{ij} t_{i-j} \, 
c^{\dagger}_{i\sigma} c^{\phantom{\dagger}}_{j\sigma}
- \mu \sum_i 
\left(c^{\dagger}_{i\sigma} c^{\phantom{\dagger}}_{i\sigma} - n 
\right) \;, \\  
H_{\rm ph} & = & \frac{1}{2 \Lambda} \sum_i \left( r^{\, 2}_i 
+ \dot{r}^2/\Omega^2 \right) \;, \\
H_{\rm el-ph} & = & \sum_i r_i 
\left(c^{\dagger}_{i\sigma} c^{\phantom{\dagger}}_{i\sigma} - n \right),
\end{eqnarray}
where we have absorbed the electron phonon coupling into the phonon
coordinate $r$ which thus has dimension of energy as does the phonon
stiffness parameter $\Lambda$. In our numerical calculations we
further specialize to a mean electron density per spin direction of $n
= 1/2$ (this implies $\mu = 0$) and we use the semicircular density of
states $\rho(\epsilon )=(1/N)\,\sum_{\vec{k}}\delta
(\epsilon-\epsilon_{ \vec{k}}) = 1/(2\pi t^2)\sqrt{4 t^{2}-\epsilon
^{2}}$ (per spin), where $\epsilon_{\vec{k}}$ is the Fourier transform
of $t_{i - j}$. However, our approach is not restricted to these
cases. The possibility of charge-density wave formation (due to a
nested Fermi surface) will not be considered in this work, but could
also be considered via our method.\\

The fundamental assumption of DMFT is the momentum independence of the
electron self energy $\Sigma(\vec{p},i\,\omega_n) \rightarrow \Sigma_n
\equiv \Sigma(i\,\omega_n)$. This implies that the physics
can be derived from an impurity model specified by the action
\begin{eqnarray}
S(\{r_{k}\};\{c_n\}) & = & (T/2\Lambda)\,
\sum_k r_{k}\,\left( 1+\omega_{k}^2/\Omega^2 \right)\,r_{k}^*\,
-n\,r_{0} 
\nonumber\\
\label{action}
& & -(2s+1)\,{\rm Tr\,\,ln}\left[ c_n\,\delta_{nm} - T\,r_{n-m} \right]\;.
\end{eqnarray}
The $c_n$ are mean field functions which describe the conduction
electrons and depend on odd Matsubara frequencies $\omega_n = 2\pi
T\,(n+1/2)$. The $r_k$ are bosonic fields which describe the phonons
and depend on even Matsubara frequencies $\omega_k = 2\pi T\,k$.  The
factor $(2s+1)$ is the spin degeneracy. The partition function may be
written as a functional integral over the bosonic fields
\begin{equation}
\label{partition}
Z = \int {\cal D}[r]\exp(-S).
\end{equation}
It is a functional of the mean field function $c$. The impurity Green
function ${\cal G}$ and self energy $\Sigma$ are defined by
\begin{equation}
\label{field}
{\cal G}_n \equiv \frac{1}{2s+1}\,\frac{\delta\,\ln Z}{\delta\,c_n}
\equiv
\frac{1}{c_n-\Sigma_n(\{c_n\})}\;.
\end{equation}
The mean field function is fixed by equating the local Green function
($G_{\rm loc}^{\rm lattice}(\omega) =
\int\,\frac{d^dk}{(2\pi)^d}\,\frac{1}{\omega-\epsilon_k-\Sigma(\omega)}
$) of the original lattice model to ${\cal G}_n$. For a semicircular
density of states one finds\cite{georges96}
\begin{equation}
\label{dmft}
c_n = i\omega _{n}+\mu -t^2\,{\cal G}_n \left( \{c_n\} \right)\;.
\end{equation}

The effective action (\ref{action}) defines an interacting quantum
field theory which cannot be solved exactly. Here we expand the action
to quadratic order in the quantum modes $\omega_k$ ($k \neq 0$) while
treating the classical modes $\omega_0$ exactly:
\begin{eqnarray}
\lefteqn{
{\rm Tr\,\,ln} \left[ c_n\,\delta_{nm} - T\,r_{n-m} \right] \approx  
{\rm Tr\,\,ln} \left[ c_n\ - T\,r_0 \right] -
}
\nonumber \hspace{0.5cm}\\
  & & \frac{T^2}{2}\sum_{n \neq m} 
(c_n - T\,r_0)^{-1}\,r_{n-m}\,(c_m - T\,r_0)^{-1}\,r_{m-n}\;.
\end{eqnarray}
The effective action $S = S_0 + \sum_{k>0} S_k$ separates into a
classical part $S_0$ and quadratic quantum parts $S_{k>0}$, both of
which depend on $r_0$:
\begin{eqnarray}
\label{class}
S_0 & = & \frac{T}{2 \Lambda}\,r_0^2 - n\,r_0
- (2s+1)\,{\rm Tr\,\,ln} \left[ c_n\ - T\,r_0 \right] \\
S_{k>0} & = & \frac{T}{\Lambda\Omega^2}\,r_k\,
\left[\omega_k^2+\Omega_k^2(r_0)\right]\,r_k^*, \label{dynac}
\end{eqnarray}
where
\begin{equation}
\Omega_k^2(r_0) = \Omega^2
\left[1+\Pi_k(r_0)\right]
\end{equation}
and
\begin{equation}
\Pi_k(r_0) = (2s+1)\,\Lambda\,T\,\sum_n 
\frac{1}{c_n-T\,r_0}\,\frac{1}{c_{n-k}-T\,r_0}\;.
\end{equation}
Since the dynamic action (\ref{dynac}) is quadratic the $k \neq 0$
modes can be integrated out exactly. The resulting partition function
\begin{equation}
Z = \int {\rm d}r\,P(r)
\end{equation}
may be written as a one-dimensional integral over the rescaled
classical coordinate $r = T\,r_0$ and
\begin{equation}
P(r) = \exp \left\{ -S_0(r)-
\sum_{k>0} \ln \left[1+\frac{\Omega_k^2(r)}{\omega_k^2}\right]
\right\} \label{Pdistr}
\end{equation}
is the (unnormalized) probability that the classical coordinate takes
the value $r$. Note that $P(r)$ depends on the mean field parameters
$c_n$ which must be computed self-consistently. Performing the
derivative with respect to $c_n$ yields the local Green function
\begin{equation}
\label{green}
{\cal G}_n = \frac{1}{Z}\,\int\,{\rm d}r\,P(r)\,
\left[\frac{1}{c_n-r} + F_n(r)\,\left(\frac{1}{c_n-r}\right)^2\right]\;.
\end{equation}
The function $F_n$ is given by
\begin{equation}
\label{full}
F_n(r) = \Lambda\,T\,
\sum_{k>0}\,\frac{\Omega^2}{w_k^2+\Omega_k^2(r)}\,
\left(\frac{1}{c_{n-k}-r}+\frac{1}{c_{n+k}-r}\right)\;.
\end{equation}
Eqs. (\ref{dmft}) and (\ref{green}) form a complete set of equations
which, in principle, can be solved for the local Green function on the
imaginary Matsubara axis. However, the direct numerical implementation
is difficult, especially if the local Green function is required for
real frequencies. In the following section we discuss physical
approximations which considerably simplify the numerical task and
allow for analytical insights.

\section{Numerical implementation}

\subsection{Static approximation}

The rapid increase of $S_{k>0}$ as the phonon frequency $\omega_k$ is
increased above the Debye frequency $\Omega$, combined with the
observation that $\Pi_k$ varies with frequency on the scale $t$, means
that when computing the $c_n$ [via Eq. (\ref{dmft})] as well as static
and energetic quantities we may replace $\Pi_k(r_0)$ by its zero
frequency limit $\Pi_0(r_0)$. (For a more formal justification see
Refs.\onlinecite{migdal58,deppeler00}.) Then $S_k$ is the action of a
harmonic oscillator with renormalized frequency $\Omega(r) \equiv
\Omega_0(r) = \Omega [1+\Pi(r,\omega=0)]^{1/2}$, and the phonon
distribution function reads
\begin{equation}
\label{semi1}
P_{\rm static}(r) = \frac{\Omega(r)/2 T}{\sinh \left(\Omega(r)/2
T\right)}\, \exp \left(-S_0(r) \right)\;.
\end{equation}
The function $F$ introduced above is given by
\begin{equation}
\label{static}
F_n(r) = \frac{\Lambda}{2}\,
\frac{\Omega^2}{\Omega(r)}\,\frac{1}{c_n-r}\left[
\coth\left(\frac{\Omega(r)}{2 T} \right) - \frac{2 T}{\Omega(r)} 
\right]\;.
\end{equation}
in the static approximation. Note that $F_n \sim 1/T$ as $T
\rightarrow \infty$.  The form (\ref{static}) is used in the iteration
of the DMFT equation (\ref{dmft}) on the Matsubara axis. We use up to
512 positive Matsubara frequencies and treat high frequencies
analytically. At the end of this first DMFT cycle we obtain the phonon
distribution function $P(r)$.\\

The scheme is stable (the `$k \neq 0$' integration exists and $P(r) >
0$) provided $\Omega^2(r) > -4\pi^2T^2$, in other words for all real
renormalized phonon frequencies and a (temperature-dependent) range of
imaginary ones. Note that $\Omega^2(r)$ increases as $|r|$ is
increased from zero. An imaginary phonon frequency corresponds to an
instability of the ground state, but if $T > |\Omega|/2\pi$ in this
formalism thermal fluctuations re-stabilize the ground state. For
imaginary $\Omega(r)$ one must analytically continue
Eqs. (\ref{semi1}), (\ref{static}) to
\begin{equation}
\label{semi2}
P_{\rm static}(r) =  \frac{|\Omega(r)|/2 T}{\sin \left(|\Omega(r)|/2
T\right)}\, \exp \left(-S_0(r) \right)
\end{equation}
and
\begin{equation}
F_n(r) =  \frac{\Lambda}{2}\,
\frac{\Omega^2}{\left|\Omega(r)\right|}\,\frac{1}{c_n-r}\left[
\frac{2 T}{\left|\Omega(r)\right|} - 
\cot\left(\frac{\left|\Omega(r)\right|}{2 T} \right)  
\right]\;,
\end{equation}
respectively. The form (\ref{semi2}) has to be used for low $T$, large
$\Lambda$, and small $r$ (for example, the case $\Lambda = 2.25$,
$\Omega = 0.1$, and $T < 0.2$ [see below]).\\

As will be discussed in more detail below an imaginary renormalized
phonon frequency is a precursor of polaronic effects. When the
temperature is lowered we have $|\Omega(r)|/2T \rightarrow \pi$. If we
introduce a small deviation $\epsilon(r)$ (increasing with $r$) via
$|\Omega(r)|/2T = \pi - \epsilon(r)$ the semiclassical phonon
distribution
\begin{equation}
P_{\rm static}(r) = \frac{\pi}{\epsilon(r)}\,\exp\left(-S_0\right)
\end{equation}
is enhanced around $r \approx 0$ in the polaronic phase. This may be
interpreted as a precursor effect of a ``polaronic band'' where
quantum oscillations allow for nearest neighbor polaron tunneling. An
analysis of the regime $T < |\Omega(r)|/2\pi$ requires a different
treatment not given here.

\subsection{Dynamic corrections}

For dynamical quantities, like the optical conductivity, frequency
corrections to $\Pi_0$ will be important at {\it low temperatures}
(they lead e.g.~to finite electron lifetimes) and the static
approximation is not sufficient. Employing the full expression
(\ref{full}) in the calculation of the local Green function is
cumbersome. To obtain ${\cal G}$ on the real frequency axis a
numerically challenging continuation from the imaginary frequency axis
has to be performed. We thus prefer to do all our calculations on the
real axis.  We begin by expanding
\begin{equation}
\label{expansion}
\Omega_k^2(r) = \Omega_0^2(r) +
\left. \partabl{\Omega_k^2(r)}{\omega_k} \right|_{k=0}\,\omega_k
+ {\cal O}(\omega_k^2)\;.
\end{equation}
On the real frequency axis this corresponds to an expansion in small
frequencies which dominate dynamic quantities at low temperatures. The
second term in Eq. (\ref{expansion}) is imaginary and leads to a
damping of the phonon modes and to small (relative order $\gamma$)
corrections to physical results.  We will neglect this feedback of the
electrons on the phonon system and keep only the frequency
renormalization, i.e. the first term in Eq. (\ref{expansion}). This
leads to
\begin{equation}
P(r)  =  P_{\rm static}(r)
\end{equation}
and
\begin{equation}
\label{expa}
F_n(r) =  \Lambda\,T\,
\sum_{k>0}\,\frac{\Omega^2}{w_k^2+\Omega^2(r)}\,
\left(\frac{1}{c_{n-k}-r}+\frac{1}{c_{n+k}-r}\right)\;.
\end{equation}
Note the difference to the static approximation [Eq. (\ref{static})]
where the finite frequency transfer of $\omega_k$ in the effective
electron fields $c_{n-k}$ and $c_{n+k}$ is missing. Eq. (\ref{expa})
defines the correct $F$ for calculating dynamic quantities from
$P_{\rm static}(r)$. In the following we will restrict ourselves to
the regime of stable quantum modes where $P_{\rm static}(r)$ is well
defined down to $T=0$. It is easy to analytically continue the
function $F$ to the real axis. The real and imaginary parts read

\begin{eqnarray}
\lefteqn{F^{'}(\omega,r) = 
\frac{\Lambda}{2} \,\frac{\Omega^2}{\Omega_r} \times} 
\nonumber \hspace{0.1cm}\\
& & 
\left[\;b(\Omega_r)\,{\cal P}^{'}(\omega+\Omega_r) -
b(-\Omega_r)\,{\cal P}^{'}(\omega-\Omega_r) - \right.
\nonumber\\
& & 
\;\;\;(2 T/\Omega_r)\,{\cal P}^{'}(\omega) + 
\nonumber\\
\label{principal}
& & 
\left.\;\;\;\frac{1}{\pi} \, {\rm P}
\int\,{\rm d}\epsilon\,{\cal P}^{''}(\epsilon)\,f(\epsilon)\,
\left(\frac{1}{\omega-\Omega_r-\epsilon} - 
\frac{1}{\omega+\Omega_r-\epsilon}\right) \right]
\end{eqnarray}
and
\begin{eqnarray}
\lefteqn{F^{''}(\omega,r) =  
\frac{\Lambda}{2}\,\frac{\Omega^2}{\Omega_r} \times}
\nonumber \hspace{0.1cm}\\
& & 
\left[\;b(\Omega_r)\,{\cal P}^{''}(\omega+\Omega_r) - 
b(-\Omega_r)\,{\cal P}^{''}(\omega-\Omega_r) - \right.
\nonumber\\
& & 
\;\;\;(2 T/\Omega_r)\,{\cal P}^{''}(\omega) -
\nonumber\\
& & 
\left.\;\;\;f(\omega-\Omega_r)\,{\cal P}^{''}(\omega-\Omega_r) +
f(\omega+\Omega_r)\,{\cal P}^{''}(\omega+\Omega_r)
\right]
\end{eqnarray}
Here $f(\omega)$ and $b(\omega)$ are the Fermi and Bose distributions,
respectively. ${\cal P}^{'}$ and ${\cal P}^{''}$ denote the real and
imaginary part of ${\cal P}(\omega+i0^+) \equiv 1/[c(\omega)-r]$,
respectively, $\Omega_r \equiv \Omega(r)$ and the integral in
Eq.~(\ref{principal}) is a principal value. If the propagators ${\cal
P}$ exhibit particle-hole symmetry then $F^{'}(\omega) =
-F^{'}(-\omega)$ and $F^{''}(\omega) = F^{''}(-\omega)$. The real part
is an odd function of frequency and the modifications arising from a
finite $\Omega_r$ may be ignored at $\omega \approx 0$. To leading
order in $\Omega_r$ we are left with
\begin{eqnarray}
\label{fun}
\lefteqn{F(\omega,r) =  \frac{\Lambda}{2}\,\frac{\Omega^2}{\Omega_r}\,
[b(\Omega_r)\,{\cal P}(\omega)-b(-\Omega_r)\,{\cal P}(\omega) +} 
\hspace{0.5cm}\\
& & 
\nonumber  
i\,f(\omega+\Omega_r)\,{\cal P}^{''}(\omega) -
i\,f(\omega-\Omega_r)\,{\cal P}^{''}(\omega)
-(2 T/\Omega_r)\,{\cal P}(\omega) ]\;. 
\end{eqnarray}
Dynamic corrections reduce the imaginary part of $F$ at small
frequencies.  For $T=0$ and $r =0$ the static approximation yields a
non-zero value $F^{''} \sim \gamma$ for small $\omega$ whereas
including finite frequency corrections give $F^{''}_{T=0} = 0$ in
a shell $-\Omega_0 < \omega < \Omega_0$.\\

The results of this section constitute an expansion in $\gamma$, and
have errors of order $\gamma^2$. However, for temperatures smaller
then the renormalized phonon frequency the (quantum modified) thermal
fluctuations become of order $\gamma^2$ and one of the order
$\gamma^2$ errors induced by the semiclassical approach leads on the
real axis to a self energy whose imaginary part changes sign, so the
self energy becomes non-causal [see below Eq. (\ref{lowtself})]. We
thus suggest two complementary {\it ad hoc\/} schemes for iterating
the DMFT equations on the real frequency axis. Both schemes result in
a causal electron self-energy and ${\cal O}(\gamma^2)$ or ${\cal
O}(\gamma T)$ corrections to physical quantities.

(i) Setting $F(\omega,r) \equiv 0$ for all $\omega$ and $r$. This
approach gives the correct high temperature behavior and is also a
good approximation for $T=0$, and we expect it to give us an idea how
the high and low temperature regimes of dynamic quantities are
connected.  Furthermore, the imaginary part of $F$ does indeed vanish
for small enough frequencies at zero temperature---as we would expect
on physical grounds since the resistivity should drop to zero in a
Fermi liquid state.

(ii) The violation of causality is only of order $\gamma^2$, as we will
show in the next section. In order to obtain a well-defined electron
self energy we may therefore add a heuristic impurity scattering rate $\tau
\sim \gamma$ to the self energy:
\begin{equation}
{\cal G} \longrightarrow
\left({\cal G}^{-1}+i\,\tau\right)^{-1}
\end{equation}
We discuss a proper choice of $\tau$ in the next section.\\ 

Note that when crossing over from the Fermi liquid to the polaronic
regime the renormalized frequency $\Omega(r=0)$ tends to zero. We
believe that for $\lambda = \lambda_c$ the `$F=0$' approach (which is
then practically identical to the classical case) becomes exact for
properties which involve only electrons at the Fermi energy. This
would imply a square root like resistivity $\rho \sim T^{1/2}$ down to
$T=0$ at this singular point.\cite{millis96} For coupling constants
slightly below $\lambda_c$ the suggested approaches (i) and (ii)
should give a reasonable picture down to lowest temperatures.\\

To summarize, the effect of quantum phonons to order $\gamma$ is to
add a semiclassical potential to the classical action. Static and
energetic quantities, including the distribution function of onsite
lattice distortions $r$, may be calculated using a simplified
(`static') semiclassical potential depending only on a renormalized
frequency $\Omega(r)$. The approach is valid for all coupling
strengths $\lambda$ but is restricted to temperatures $T >
|\Omega(r=0)|/2\pi$ in the polaronic regime $\lambda > \lambda_c
\approx 2.2$ where $\Omega(r)$ becomes imaginary. Some additional
numerical effort (compared to the classical case\cite{millis96}) is
required due to the calculation of the renormalized frequencies. Care
has to be taken when calculating dynamical quantities at low
temperatures due to the presence of an additional quantum scattering
term. We propose two schemes for incorporating dynamic corrections on
top of the simplified semiclassical action. For coupling strengths
$\lambda < \lambda_c$, i.e. in the Fermi liquid regime, the crossover
between the high and low temperature behavior of transport properties
can be studied.

\section{Discussion and Observables}

\subsection{Fermi liquid to polaron crossover}

When discussing the influence of local lattice fluctuations on
electronic properties one has to distinguish between two regimes. For
small values of $\Lambda$ the conduction electrons are weakly
renormalized quasiparticles. This is the Fermi liquid regime. For
large $\Lambda$ the conduction electrons are self-trapped by their own
local lattice distortions. This is the polaronic regime. In the
classical case\cite{millis96} ($\gamma = 0$) the main features of the
``polaronic transition'' may be illustrated by expanding to quadratic
order around $r = 0$:
\begin{equation}
\label{classlo}
S_0 = \frac{1}{2\,\Lambda\,T} 
\left(1+\Lambda\,\Gamma_2 \right)\,r^2
\end{equation}
with 
\begin{equation}
\label{Gamma}
\Gamma_{\alpha} = (2s+1)\,T\,\sum_n\,c_n^{-\alpha} \;. 
\end{equation}
For later use we define
\begin{equation}
\bar{\Lambda}_{\rm classical} =  \frac{\Lambda}{1+\Lambda\,\Gamma_2},
\;\;\;\;
\bar{\Omega}_{\rm classical} = \Omega\,\sqrt{1+\Lambda\,\Gamma_2}\;.
\end{equation}
As long as $\Lambda < -1/\Gamma_2 \equiv \Lambda_c^{\rm classical}$,
$S_0$ is minimized by $r = 0$. In the $T \rightarrow 0$ limit the
dominant contributions to $Z$ come from small distortions. If
$\Lambda$ approaches $\Lambda_c^{\rm classical}$ from below, the
elastic energy vanishes and the system enters a polaronic state with
$\langle r \rangle \neq 0$. The critical $\Lambda_c$ depends on band
filling, temperature, $\Lambda$, $\Omega$, as well as other parameters
(such as electron-electron interaction strength). In the
noninteracting case [where $ c^0_n = \frac{i}{2}\,[\omega_n +
\sqrt{\omega_n^2+4 t^2} ]$ (upper half plane)] and at $T = 0$ we find
$\Lambda_c = 3 \pi t/[ (2s+1)\,4]$. It has been shown\cite{millis96}
that at high temperatures electrons are strongly affected by the
presence of a local (polaronic) lattice instability. A strongly
coupled system turns from a Fermi liquid into a polaronic insulator
with activated conduction caused by the hopping of small polarons. In
the crossover regime the system can be characterized as a ``bad
metal''. This intermediate state is the focus of the current paper.

We now discuss the changes to this picture introduced by quantum
effects. We first focus on the regime of stable quantum modes,
i.e. $\Lambda < \Lambda_c$.  For small lattice distortions we may
expand the effective phonon frequency around $r = 0$.  At half filling
we find
\begin{equation}
\Omega^2(r) = \Omega^2\,
\left(1+\Lambda\,\Gamma_2+3\,\Lambda\Gamma_4\,r^2\right)\;.
\end{equation}
In the regime under consideration we can assume $1+\Lambda\,\Gamma_2 >
0$ and use Eq. (\ref{semi1}) to calculate the effective partition
function. To second order in $r$ and at low temperatures
the semiclassical action reads
\begin{eqnarray}
\lefteqn{S =  \frac{1}{2\,\Lambda\,T} \times}
\nonumber \hspace{0.3cm}\\
& & 
\left[ \Lambda\,\Omega\,\sqrt{1+\Lambda\,\Gamma_2} + \left(
1+\Lambda\,\Gamma_2 + 
\frac{3}{2}\,\frac{\Lambda^2\,\Omega}{\sqrt{1+\Lambda\,\Gamma_2}}\,
\Gamma_4 \right)\,r^2\right] \nonumber\\ 
& & 
\label{semilo}
\equiv S(r=0) + \frac{1}{2\,\bar{\Lambda}\,T}\,r^2\;.
\end{eqnarray}
The first ($r=0$) term is the Migdal correction to the classical
action given in Eq. (\ref{classlo}). It arises from the one-phonon-loop
diagrams.\cite{deppeler00} Quadratic and higher orders in the
distortions have to be considered at higher temperatures or stronger
couplings, especially if the system is close to the instability $\Lambda_c$.
In terms of the dimensionless parameters
\begin{equation} 
\lambda = \Lambda/t \, , \;\;\; \gamma = \Omega/t \, , 
\end{equation} 
the leading corrections to the classical action are of the form
$\bar{\lambda}^n_{\rm classical}\,\bar{\gamma}_{\rm classical}$ with
$n \geq 2$, i.e. they are of order $\bar{\gamma}_{\rm classical}$.
Notice however that the corrections can become arbitrary large if
$\Lambda \rightarrow \Lambda_c^{\rm classic}$.\cite{deppeler01}

The $T=0$ polaronic instability occurs when the coefficient of $r^2$
in Eq. (\ref{semilo}) vanishes, i.e.~when (in rescaled units)
\begin{equation}
1 + \lambda_c\,(t\Gamma_2)+
\frac{\lambda_c^2\,\gamma\,3(t^3\Gamma_4)/2}{\sqrt{1+\lambda_c\,(t\Gamma_2)}}
 =  0 \;.
\end{equation}
This equation has a solution at $\lambda_c = \lambda_c(\gamma) =
-\Gamma_2/t + {\cal O}(\gamma^{2/3})$. Because $\Gamma_4 > 0$ (for the
Holstein model at half filling $\Gamma_4 = (2s+1)\,8/(15\,\pi\,t^3)$
in the noninteracting limit) we see that quantum fluctuations increase
the critical coupling needed for the polaronic instability by an
amount $\delta\lambda \sim \gamma^{2/3}$. Note also that the sign of
$\bar{\Omega}^2_{\rm classical} = \Omega^2(r=0)$ is only determined by
$\Gamma_2$. The system enters the regime of unstable quantum modes at
coupling strength $\lambda \approx \lambda_c^{\rm classic} <
\lambda_c(\gamma)$. As we shall see, this leads to an interesting
precursor effect of polaronic behavior.

\subsection{Low-temperature self energy}

In this section we discuss the low temperature self energy in detail,
in order to clarify the limitations of our theory. For weak to
intermediate coupling strengths, $\Lambda < \Lambda_c$, $P(r)$ is
peaked about $r = 0$ with half-width of order $T$, so an expansion in
$r$ is justified if $\bar{\lambda}\,T/t \ll 1$, i.e.~if $T$ is low and
the system is not too close to the polaronic instability. It is
convenient to rewrite Eq. (\ref{green}) to order ${\cal
O}(\bar{\lambda}_{\rm classical}^2\,\bar{\gamma}_{\rm classical}^2,
\bar{\lambda}_{\rm classical}^2\,T^2/t^2)$ by adding and subtracting
the $k=0$ term in $F_n$ and moving all terms to the denominator, so
(denoting $\frac{1}{Z}\,\int\,{\rm d}r\,P(r)$ by $\langle\,\rangle$)
\begin{equation}
{\cal G}_n = \left\langle
\frac{1}{c_n-r-\Sigma_n^{\rm Migdal}(r)
-\bar{\Lambda}_{\rm classical}(r)\,T/(c_n-r)}
\right\rangle
\end{equation}
with 
\begin{equation}
\Sigma_n^{\rm Migdal}(r) = \Lambda\,T
\sum_{k=-\infty}^{k=\infty} 
\frac{\Omega^2}{\omega_k^2+\Omega_k^2(r)}\,
\frac{1}{c_{n-k}-r}\;.
\end{equation}
At low $T$ and not too close to the polaronic instability we may
proceed by expanding ${\cal G}$ to order $r^2$, averaging, and then
replacing terms in the denominator. This leads to a self energy
\begin{equation}
\label{lowtself}
\Sigma_n^{\rm low T} = \Sigma_n^{\rm Migdal}(r=0)
+\bar{\gamma}_{\rm classical}\,T\,A_n\;. 
\end{equation}
The expression for $A_n$ is somewhat cumbersome, because many terms
contribute and the result depends on the relative values of external
and phonon frequency. However, in all cases we have examed, the
analytic continuation of $A_n \rightarrow A(\omega+i0^+)$ has an
imaginary part of positive sign, corresponding to a non-causal
contribution to $\Sigma$, and magnitude of order unity
($\bar{\gamma}^0$). For example, if one assumes that all relevant
contributions in the occuring Matsubara sums stem from small
frequencies $|\omega_{n\pm k}| << 2t$ then (noting $c_n = i\,t\,{\rm
sgn}\,\omega_n$ at half filling)
\begin{eqnarray}
\label{coeff}
A(\omega = 0) & = & i\,
\frac{\bar{\lambda}_{\rm classical}-\bar{\lambda}}{\bar{\gamma}_{\rm classical}}
+i\,\bar{\lambda}_{\rm classic}\,\bar{\lambda} \nonumber \\
& = &
i\,\left[
\frac{3}{2}\,\bar{\Lambda}^3_{\rm classic}\,\Gamma_4 +
\bar{\lambda}_{\rm classic}\,\bar{\lambda}
\right]\;.
\end{eqnarray}
The non-causal self energy is of order ${\cal O}(\bar{\gamma}_{\rm
classical}\,T)$, i.e.~formally ${\cal O}(\bar{\gamma}^2_{\rm
classical})$, and is seen from the derivation to arise from an
incomplete treatment of the ${\cal O}(\gamma^2)$ contributions to
physical quantities. Unfortunately at $(\omega,T) \ll \Omega$, the
leading contribution $\Sigma_{\rm Migdal} \sim \Omega\,s(\Omega/T)$
where $s \sim \exp[-\Omega/T]$ as $T \rightarrow 0$, so the non-causal
correction term dominates.

This may be viewed in a different way: a purely classical
approximation ($\Omega \rightarrow 0$) would lead to a $T$-linear
scattering rate with coefficient of order unity. The semiclassical
approximation overcorrects for this behavior, cancelling the leading
term leaving a correction of order $\gamma T$ but of the wrong
sign.

The unphysical low temperature behavior is not due to the realization
of the semiclassical approach introduced in Eq. (\ref{expa}). Starting
from the full expressions Eqs. (\ref{green}) and (\ref{full}) we find
a similar result
\begin{eqnarray}
\lefteqn{
\bar{\Lambda}_{\rm classic}- \bar{\Lambda} \approx
\bar{\Lambda}^2_{\rm classic}\,\Lambda\,2(2s+1)\,T^2 \times}
\nonumber \hspace{0.5cm}\\
& & 
\sum_n\,\sum_{k>0}
\frac{\Omega^2}{\omega_k^2+\Omega_k^2}\,
\left[
\frac{1}{c^3_n}\left(\frac{1}{c_{n+k}}+\frac{1}{c_{n-k}}\right) +
\frac{1}{c_n^2}\frac{1}{c_{n+k}^2}
\right] 
\end{eqnarray}
which will also lead to a non-causal self energy. We conclude, that
higher orders of quantum contributions $\gamma^n$ are required to
completely suppress the thermal contributions at low temperatures $T <
\bar{\Omega}$ (which wrongly would lead to a linear resistivity in
$T$).

However, the thermal fluctuations are reduced in the semiclassical
approach compared to the classical ($\gamma = 0$) case by a factor
$\gamma$ and the low-temperature self-energy is of order
$\bar{\gamma}^2_{\rm classic}$, as can be seen from Eqs. (\ref{lowtself})
and (\ref{coeff}):
\begin{equation}
\Sigma^{''}(\omega=0) = \frac{\bar{\Omega}^2_{\rm classic}}{t}\,
f\left(\frac{T}{\bar{\Omega}_{\rm classic}}\right)\;,
\end{equation}
where $f$ is a scaling function.  Therefore, an order $\bar{\gamma}_{\rm
classic}$ impurity scattering rate $\tau$, as proposed in the
previous section, will lead to a well-defined electron self-energy. An
appropriate choice may be found from Eq. (\ref{static}).  In the
classical limit ($\bar{\Omega}_{\rm classic} \rightarrow 0$) we find
\begin{equation}
F_n(r=0) = \frac{\Lambda}{12}\,\frac{\Omega^2}{T}\,c_n^{-1},
\end{equation}
i.e. a scattering rate $\tau \approx \bar{\lambda}_{\rm
classical}\,\bar{\Omega}_{\rm classical}/12$ for temperatures $T \leq
\bar{\Omega}_{\rm classical}$. This is the value we choose for the
additional impurity contribution. Note that in the quantum limit ($T
\rightarrow 0$) the static approach gives $\tau \approx
\bar{\lambda}_{\rm classical}\,\bar{\Omega}_{\rm classical}/2$.

\subsection{Observables}

We end this section by introducing the four observables we have
investigated numerically: phonon distribution function, electron
spectral function, optical conductivity, and resistivity. The
normalized {\em phonon distribution function\/} $p(r) = P(r)/Z$ was
defined above.  Within DMFT the spectral function and optical
conductivity can be derived from the electron self-energy on the real
frequency axis. The {\it density of states} is given by
\begin{equation}
\rho(\omega) =  \int {\rm d}\epsilon\,
\rho_0(\epsilon)\,\rho(\epsilon,\omega) 
\end{equation}
with the spectral function defined by 
\begin{equation}
\rho \left( \epsilon_{\vec{k}},\omega \right)  =  
-\frac{1}{\pi}\,{\rm Im} 
\left[
\frac{1}{\omega+\mu-\Sigma(\omega)-\epsilon_{\vec{k}}}
\right]\;.
\end{equation}
The {\it optical conductivity} is obtained from linear response
theory applied to the expectation value of the current operator
\begin{equation}
j = \sum_p \left(\partabl{\epsilon_p}{p}\right)\,c_p^{\dagger}c_p
\end{equation}
In performing the sums over momentum $p$ we will assume a hypercubic
lattice.\cite{pruschke95} The differences to a Bethe lattice are not
significant for the results presented here. The locality of the
electron self energy implies that the momentum summation can be
performed independently on each side of an electron-phonon vertex when
calculating the expectation value of the current. Since the electron
velocity $\partial \epsilon_p / \partial p$ is an odd function of
momentum all vertex corrections vanish and the optical conductivity is
obtained by a convolution of two full Green functions:

\begin{eqnarray}
\lefteqn{\sigma(\Omega) = \sigma_0\times}
\nonumber \hspace{0.1cm} \\
& & 
\label{cond}
\int {\rm d}\omega\,
\left(-\frac{f(\omega+\Omega)-f(\omega)}{\Omega}\right)\,
\int {\rm d}\epsilon\,
\rho_0(\epsilon)\,\rho(\epsilon,\omega)\,\rho(\epsilon,\omega+\Omega)
\end{eqnarray}

where $\sigma_0$ is a constant and $f(\omega)$ is the Fermi
distribution. The {\it resistivity} of the system can be deduced from
the optical conductivity via $\rho_{\rm DC} = 1/\sigma(0)$.  We
briefly discuss $\rho_{\rm DC}$ in the regime $\Lambda < \Lambda_c^{\rm
classic}$ where the quantum modes are stable. If
$\Sigma^{''}(\omega=0) \ll 2t$ as $T \rightarrow 0$, the low
temperature DC conductivity is given by
\begin{equation}
\sigma(0) = \frac{\pi}{2}\,\sigma_0\,\rho(0)\,
\int\,\frac{{\rm d}\omega}{2 T}\,\frac{1}{\cosh^2(\omega/2T)}\,
\frac{1}{|\Sigma^{''}(\omega,T)|}
\end{equation}
and we expect (for a Fermi liquid ground state) $\rho_{\rm DC} \sim
{\rm const}+T^2$. However, due to the low-temperature defect of the
semiclassical approach discussed above this regime cannot be
reached.

\section{Results}

In this section we apply the semiclassical method developed above to a
system of spinless electrons ($s = 0$) interacting with local lattice
distortions, which is relevant to half-metals such as CMR manganites.
First we consider {\it high to intermediate
temperatures}. As discussed in section III.B the high to low
temperature crossover of dynamical quantities
can be discussed within the `$F = 0$'
implementation of the semiclassical approach. This will be done first.
A detailed comparison with other approaches
will be given below, when discussing the low temperature behavior.\\

\begin{figure}
\centerline{\psfig{file=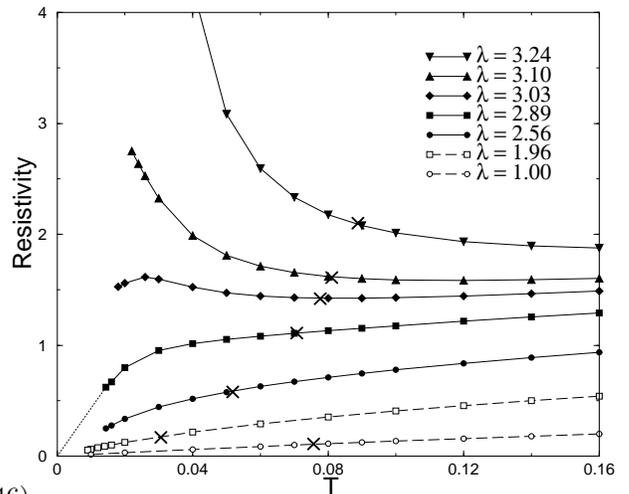,width=8cm,angle=-90}}
\vspace*{0.1cm}
\caption{\label{res.fig} \footnotesize Resistivity as a function of
temperature $T$ for various electron-phonon coupling strengths
$\lambda$ at a given phonon frequency $\gamma = 0.1$. Temperatures
corresponding to the renormalized phonon frequency $|\Omega(r=0)|$
(taken at $T = 0.2$) are indicated by a cross for each value of
$\lambda$. Dashed and solid lines correspond to the regime of stable
($\lambda < \lambda_c^{\rm class}$) and unstable ($\lambda >
\lambda_c^{\rm class}$) quantum modes, respectively. The low $T$
behavior in the regime of unstable quantum modes is beyond the scope
of this paper and the dotted line for $\lambda = 2.89$ serves as guide
to the eye.}
\end{figure}
Fig. \ref{res.fig} shows the resistivity as a function of temperature
for a wide range of coupling strengths and $\gamma = 0.1$. For narrow
electron bands (typical for e.g.~the A-15 materials) this choice
corresponds to setting $\Omega \sim 300$K. At high temperatures $T \gg
\Omega$ we reproduce the results of the classical ($\gamma = 0$)
approach.\cite{millis96,millis99} At weak couplings the resistivity
depends linearly on temperature $\rho(T) = AT + B$. At intermediate
couplings the resistivity becomes nonlinear in $T$ and the ratio $A/B$
decreases.  As discussed in detail in Ref. \onlinecite{millis99} this
is essential for the phenomenon of `resistivity saturation' in models
which couple local lattice vibrations to the level positions: the
scattering rate of the electrons with lattice fluctuations increases
with $T$, but less rapidly than predicted from second order
perturbation theory. Notice that when discussing resistivity saturation
we focus on temperatures above the (renormalized) Debye temperature
indicated by a cross in Fig. \ref{res.fig}. As seen in
Fig. \ref{res.fig} the system remains metallic for coupling strengths
well above $\lambda_c^{\rm classic} = 2.36$ because quantum lattice
fluctuations allow the electrons to tunnel between neighboring
sites. The crossover to the high temperature resistivity (which is
large) is more pronounced than for a conventional metal. We thus refer
to this state as a {\it bad metal}. Our approach is capable of
describing this interesting state where quantum (Migdal) and thermal
(classical) fluctuations compete. The strong renormalization of the
phonon frequencies leads to a drop in resitivity at a temperature
which depends strongly on $\lambda$ and which is well below the
unrenormalized Debye temperature $\Omega = 0.1$.  Only for stronger
couplings $\lambda > 3.03$ does the resistivity show insulating
behavior over the entire accessible temperature range.\\

\begin{figure}
\centerline{\psfig{file=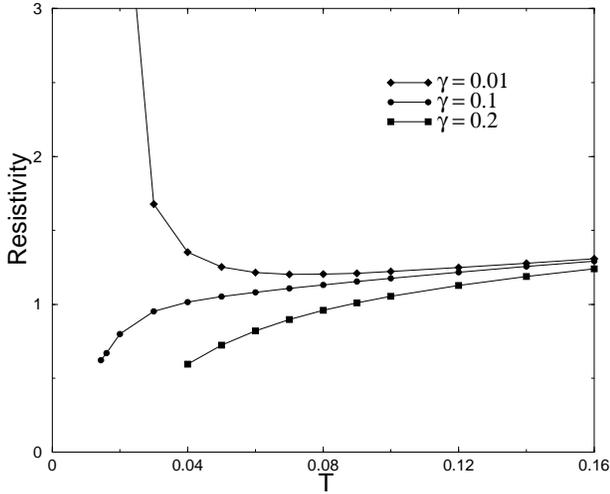,width=8cm,angle=-90}}
\vspace*{0.1cm}
\caption{\label{res2.fig} \footnotesize Resistivity as a
function of temperature $T$ for intermediate electron-phonon coupling
strength $\lambda = 2.89 > \lambda_c^{\rm class} = 2.36$ and various
phonon frequencies. Quantum fluctuations destroy the
polaronic insulator and restore a metallic state.}
\end{figure}
The new feature brought by quantum fluctuations to these curves is a
coexistence of very weak temperature dependence at high $T$ with
metallic behavior at low $T$. This is clearly shown in
Fig. \ref{res2.fig}, which displays the effect of changing quantum
fluctuations on the $\lambda = 2.89$ curve of Fig. \ref{res.fig}. One
sees that the nearly classical model ($\gamma = 0.01$) is an insulator
at low $T$, while the other two are metals, while the higher-$T$
behavior is hardly affected. The $\gamma = 0.1$ data bear a striking
resemblance to the resistivity of well known saturation materials such
as $\rm Nb_3Sb$.\cite{fisk76} We also note that the data shown in
Fig. \ref{res2.fig} imply an {\it isotope-driven metal insulator
transition}, although to obtain this coupling one must finetune
$\lambda$ within ${\cal O}(\gamma)$ of the critical value.\\

\begin{figure}
\centerline{\psfig{file=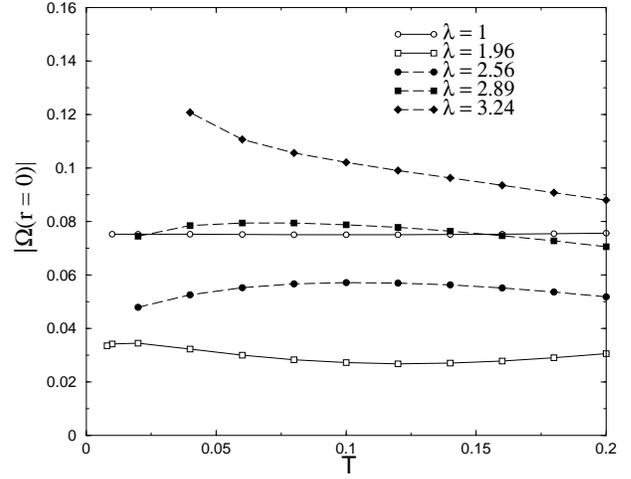,width=8cm,angle=-90}}
\vspace*{0.1cm}
\caption{\label{omega.fig} \footnotesize Renormalized phonon frequency
$|\Omega(r = 0)|$ as function of temperature $T$ for various coupling
strengths $\lambda$ at a given phonon frequency $\gamma = 0.1$. Close
to the (classic) polaronic instability a clear temperature dependence
is visible. For $\lambda = 1.96$ the relative change of $|\Omega(r =
0)|$ amounts to $12\%$ over the plotted temperature regime. Solid and
dashed lines indicate the regime of stable [real $\Omega(r = 0)$] or
unstable [imaginary $\Omega(r = 0)$] quantum modes, respectively.}
\end{figure}
In Fig. \ref{res.fig} we indicated the modulus of the renormalized
phonon frequency by a cross. However, $|\Omega(r = 0)|$ may be
considerably temperature dependent. Fig. \ref{omega.fig} shows the
temperature dependence of the renormalized phonon frequency $|\Omega(r
= 0)|$. For weak couplings (e.g. $\lambda = 1$) the frequency is
slightly suppressed and nearly independent of temperature. However,
close to the (classic) polaronic instability $\lambda_c^{\rm classic}$
(e.g. $\lambda = 1.96$) the phonon frequency is strongly renormalized
towards zero and temperature dependent. Nonmonotonic temperature
dependence of phonon frequencies was found in models with interactions
between phonons and strongly correlated electrons and was used to
interpret anomalies in Raman scattering and acoustic experiments on
certain superconducting molecular crystals.\cite{merino00} Here we
show that even pure electron-phonon coupling may lead to a
nonmonotonic temperature dependence of the phonon frequency. Note,
however, that the connection between $|\Omega(r = 0)|$ and measurable
quantities remains yet to be established. On entering the regime of
unstable quantum modes the temperature dependence of the renormalized
phonon frequency change qualitatively. In addition, an enhancement of
the phonon frequency is possible. In the insulating regime $|\Omega(r
=0)|(T)$ increases monotonically.\\

\begin{figure}
\centerline{\psfig{file=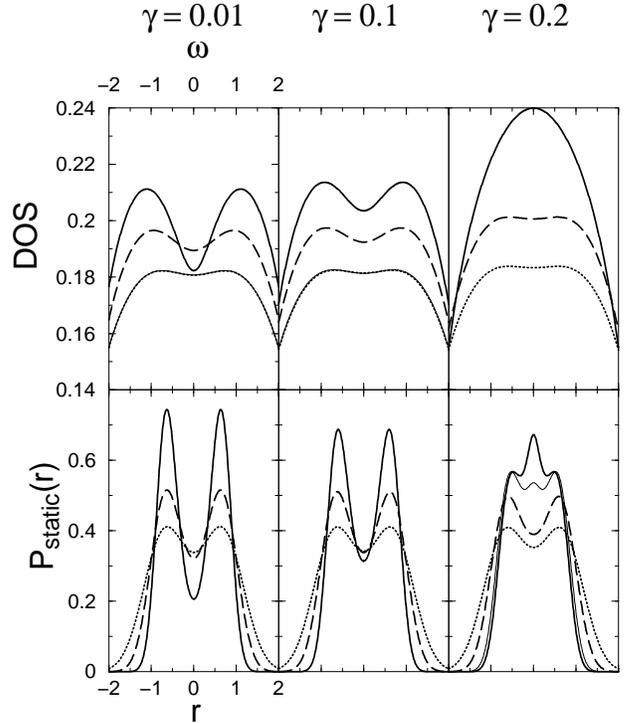,width=8cm,angle=-90}}
\vspace*{0.1cm}
\caption{\label{disdos.fig} \footnotesize LOWER PANEL: Distribution of
local distortions $P_{\rm static}(r)$ for three different phonon
frequencies and $T = 0.2$ (dotted lines), $T = 0.1$ (dashed lines),
and $T = 0.04$ (solid lines). The coupling strength equals $\lambda =
2.89$. Scales on $x$- and $y$-axes are the same in all plots and given
in the first one. UPPER PANEL: Electron spectral function as function
of frequency. The parameter values are identical to those in the lower
panel. Scales on $x$- and $y$-axes are the same in all plots and given
in the first one.}
\end{figure}
\begin{figure}
\centerline{\psfig{file=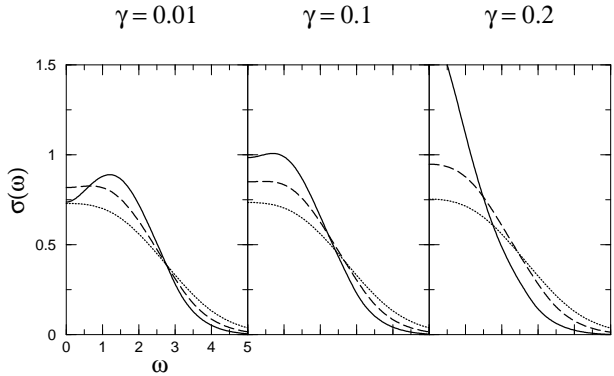,width=8cm,angle=-90}}
\vspace*{0.1cm}
\caption{\label{sig.fig} \footnotesize Optical conductivity
$\sigma(\omega)$ for three different phonon frequencies and $T = 0.2$
(dotted lines), $T = 0.1$ (dashed lines) and $T = 0.04$ (solid
lines). The coupling strength equals $\lambda = 2.89$. Scales on $x$-
and $y$-axes are the same in all plots and given in the first one.}
\end{figure}

The collapse of the polaronic state due to quantum fluctuations causes
other changes in electronic properties which are summarized in
Figs. \ref{disdos.fig} and \ref{sig.fig}. The lower panel of
Fig. \ref{disdos.fig} shows the distribution of local lattice
distortions for various frequencies and the uppper one the spectral
function. Fig. \ref{sig.fig} displays the optical conductivity. At
high temperatures we find a considerable probability for finite
(dynamic) lattice distortions even in the metallic state. The density
of states at the Fermi energy is relatively small and the Drude peak
in the optical conductivity is very broad. In fact, the optical
spectral weight can be spread over a frequency range which continously
increase with increasing temperature. Therefore the term
``saturation'' is a misnomer: There is no intrinsic maximum value of
the high-temperature resistivity.\cite{millis99,calandra01}. Upon
lowering the temperature we recover a conventional metallic state,
where the probability of strong lattice distortions decreases, the
density of states at the Fermi energy increases, and the Drude peak
sharpens.\\

In a polaronic insulator there are pronounced negative (positive)
distortions indicating that the electrons (holes) remain on a given
site for a sufficiently long time for the lattice to relax. If the
phonon frequency increases the probability at $r \approx 0$ becomes
larger, which points to an increased electron mobility. As a
consequence, three maxima in $P_{\rm static}(r)$ may be observed when
crossing over from high to low temperatures for sufficiently large
phonon frequency. The upper panel of Fig. \ref{disdos.fig} shows the
corresponding changes in the spectral function of the electrons. The
polaronic insulator exhibits a (pseudo) gap. The `splitting' of the
band is caused by dynamical local distortions. Upon increasing
$\gamma$ one sees that at low $T$ considerable spectral weight is
transferred from high to low energies, filling up the gap, whereas the
spectral function at high $T$ is virtually unchanged. Further insight
may be obtained by considering the optical conductivity
(Fig. \ref{sig.fig}). For $\gamma = 0.01$ we see an insulating gap
develop from a broad, incoherent high-$T$ conductivity; again
increasing the phonon frequency leads to the development of a low $T$
metallic state.\\

\begin{figure}
\centerline{\psfig{file=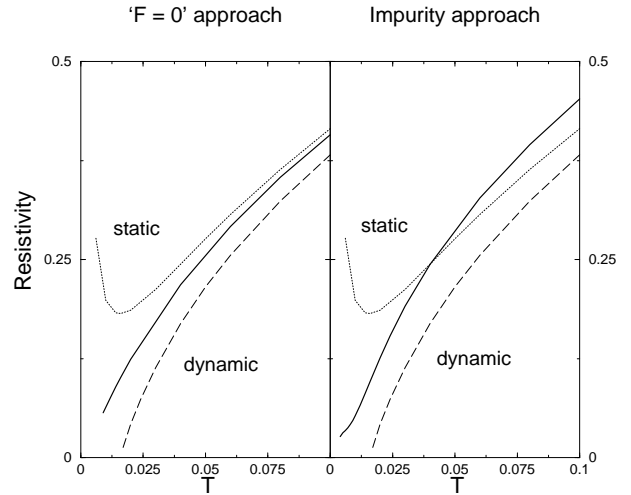,width=8cm,angle=-90}}
\vspace*{0.1cm}
\caption{\label{resvgl.fig} \footnotesize Resistivity as a function of
temperature $T$ for $\lambda = 1.96$, $\gamma = 0.1$, calculated using
various implementations of the semiclassical approach: (i) Static
approximation $\Pi_k \rightarrow \Pi_0$ (static, dotted line), (ii)
expansion in $\omega_k$ (dynamic, dashed line), (iii) neglecting $F$
(solid line, left panel) and (iv) adding an impurity scattering (solid
line, right panel). For details see text and section III. The
deficiencies of approaches (i) (nonmonotonic resisitivity) and (ii)
(negative resisitivity, i.e.~non-causal self energy) are clearly
visible.}
\end{figure}
The technical difficulties mentioned in the previous section limited
the calculation shown in Fig. \ref{disdos.fig} and \ref{sig.fig} to
relatively high temperatures. We now turn to the {\it
low-temperature\/} behavior ($T < \Omega$) to reveal the strengths and
weakness of different low-$T$ implementations of the semiclassical
approach. We focus on $(\lambda,\gamma) =
(1.96,0.1)$. Fig. \ref{resvgl.fig} shows the low-temperature
resistivity, calculated using various implementations of the
semiclassical approach. Quantum lattice fluctuations enter in two
places: the distribution function of lattice distortions $P(r)$ and
the generalized contribution $F_n(r)$ to the electron self energy. The
two properties influence the resistivity in opposite ways. A higher
probability of small distortions decreases the resistivity whereas a
negative $F''(\omega,r)$ describes additional electron-lattice
scattering.\\

The implementations discussed here only differ in the choice of $F$ as
discussed in section III.B. The static approximation gives $F_n(r=0) =
(\bar{\Lambda}_{\rm classic}\,\bar{\Omega}_{\rm classic}/2)\,
c_n^{-1}-\bar{\Lambda}_{\rm classic}\,T\,c_n^{-1}$ at very low
temperatures and leads to a nonmonotonic resistivity, as seen in
Fig. \ref{resvgl.fig} (dotted line). When we take into account that
$F$ has to be reduced for energies around the Fermi energy, the finite
resistivity at zero temperature of the static approximation is
eliminated but the resulting electron self energy now violates
causality. Fig. \ref{resvgl.fig} shows that the resistivity drops to
zero at a finite temperature (dashed line). We conclude that at very
low temperatures dynamic corrections of higher orders (see section
III.B) become important.  To mimic their effect we suggested, first,
to set $F \equiv 0$. This is good at high to intermediate temperatures
but cannot account for a Fermi liquid like resisitivity, $\rho_{\rm
DC} \sim {\rm const}+T^2$ as seen in the left panel of
Fig. \ref{resvgl.fig} (solid line). Second, we suggested adding an
impurity scattering of order $\gamma$ to obtain a well defined
electron self energy. This is shown in the right panel of
Fig. \ref{resvgl.fig} (solid line): at very low temperatures the
resistivity levels off to a finite value due to the added impurity
scattering. Both (`$F=0$' and impurity) approaches demonstrate the
development of a resistivity $\rho \sim T^{1/2}$ at $\lambda =
\lambda_c^{\rm classical}$ and $T \rightarrow 0$.\\

\begin{figure}
\centerline{\psfig{file=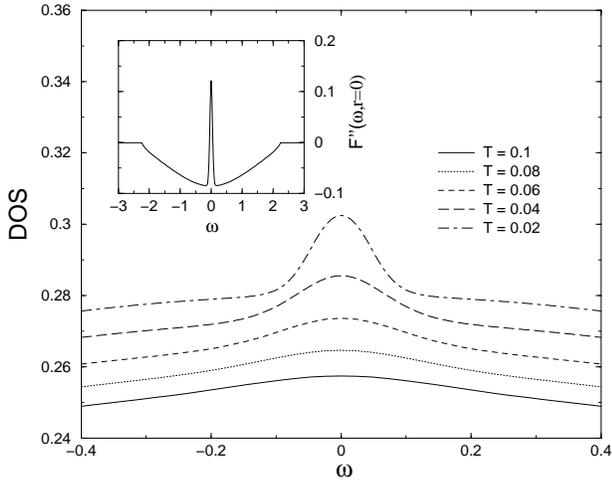,width=8cm,angle=-90}}
\vspace*{0.1cm}
\caption{\label{lowdos.fig} MAIN PANEL: Density of states as function
of frequency near the Fermi energy for $(\lambda,\gamma)$ =
$(1.96,0.1)$ and various temperatures. The polaronic feature which
appears at $\omega \approx 0$ is responsible for the strong decrease
of the resistivity. INSET: Imaginary part of the generalized
contribution of the quantum lattice fluctuations to the electron self
energy $F(\omega,r)$ as function of frequency $\omega$ at zero
distortion $r=0$ (parameter as in the main panel and $T=0.02$). Energy
conservation reduce the scattering in a shell around the Fermi energy.
}
\end{figure}
The inclusion of impurity scattering is an {\it ad hoc\/}
remedy. However, it can be used to demonstrate the formation of
polaronic bands.  In the calculation leading to the spectral function
in Fig. \ref{disdos.fig} we neglected the generalized contributions of
the quantum lattice fluctuations to the electron self energy
completely, i.e.~$F \equiv 0$. In a bad metal, as seen in
Fig. \ref{disdos.fig}, this leads to an increase of spectral weight in
a broad energy range around the Fermi energy when the temperature is
lowered. However, $F$ should be frequency dependent. Additional
scattering of electrons with quantum lattice fluctuations is possible
for electrons sufficiently far from the Fermi energy as enforced by
energy conservation (absorption and emission of finite frequency
phonons). The frequency dependence of the scattering rate is
illustrated in the inset of Fig. \ref{lowdos.fig} which displays the
imaginary part of $F(\omega,r=0)$ as given in Eq. (\ref{fun}). Even
the sign changes near $\omega = 0$ which compensates partially
scattering from thermal fluctuations. This physics is captured by the
impurity approach and leads to a clearly visible polaronic band in the
spectral function as seen in the main panel of
Fig. \ref{lowdos.fig}. For weak to intermediate coupling strengths,
$\Lambda < \Lambda_c^{\rm classic}$, the phonon distribution function
$P_{\rm static}(r)$ is concentrated around $r \approx 0$. Therefore,
at $T=0$ the electron self energy is given by the Migdal expression,
see Eq. (\ref{lowtself}), and a similar polaronic peak in the spectral
function like here at finite temperatures can be observed at zero
temperature.\cite{hague01} However, the three peak structure of $P(r)$
as seen in Fig. \ref{disdos.fig} developing close to
$\Lambda_c(\gamma)$ is clearly beyond the Migdal approach.\\

\section{Conclusion}

In this paper we have developed a semiclassical approach based on the
dynamical mean field theory to treat the interactions of electrons
with local dynamic lattice distortions. The method is not restricted
to small or strong electron-phonon couplings. It can be applied to
three spatial dimensions and may be extended to include interactions
other than electron-phonon. The effective action is organized in terms
of local static $n$-point correlation functions $\Gamma_n$ which have
to be appropriately modified in the presence of other interactions.
However, the frequency dependence of $\Gamma_n$ cannot be neglected
alltogether. At low temperatures it leads to additional
electron-lattice scattering which is important when calculating
dynamical quantities. We discussed the effect of these dynamic
corrections extensively. Nevertheless, errors of order $\gamma^2$
inherent in the semiclassical approach allow only for a qualitative
discussion of the low temperature behavior by introducing an impurity
scattering of order $\gamma$ by hand. A quantitative analysis is
beyond the scope of this paper and left for future work. However, at a
critical coupling $\lambda = \lambda_c^{\rm classic} = 3\pi/[(2s+1)4]$
a classical picture should hold and the resistivity becomes $\rho \sim
T^{1/2}$ for very low temperatures.\\

We have applied our method to study isotope effects on electronic
properties in the crossover regime between Fermi liquid and polaronic
behavior.  We found large effects including an isotope-driven
insulator-to-metal transition. The phenomena described in Sec. V may
be relevant for various half-metals such as CMR manganites and A15
compounds even though the simplicity of the model studied here does
not allow for detailed comparisons with experiments. Nevertheless, an
isotope-driven insulator-to-metal transition was
observed\cite{belova00} in $\rm
(La_{0.25}Pr_{0.75})_{0.7}Ca_{0.3}MnO_3$. Within the picture proposed
above this can be understood as a quantum tunneling effect.  In
addition to electron-phonon interactions, a realistic description of
conduction electrons in the CMR manganites must account for double
exchange with the core spins of the Mn ions. This would amplify the
insulator-to-metal transition observed above: if the electrons become
mobile the core spins tend to order ferromagnetically in order to
further lower the electron kinetic energy.\\

We have also addressed the question of resistivity saturation in
metals with high resistivity. Quantum lattice fluctuations shift the
polaronic instability to larger couplings $\delta\lambda \sim
\gamma^{2/3}$. Therefore, systems with strong quantum lattice
fluctuations remain metallic up to larger electron-phonon couplings
than systems with purely classical (thermal) fluctuations which
implies a stronger violation of the validity of second order
perturbation theory.  We may (misleadingly) speak of resistivity
saturation although there is no upper bound for the high-temperature
resistivity in contrast to models which couple atomic vibrations to
electron hopping matrix elements.\cite{calandra01} The resistivity at
high temperatures assumes very large values and a pronounced change in
slope of the temperature dependent resitivity is obtained when
crossing over from high to low temperatures. The smoothness of the
crossover can be changed by isotope replacements without changing the
strength of the electron-phonon coupling.\\

{\it Acknowledgements} SB thanks M.~Calandra and G.~Zwicknagl for
useful discussions.  SB acknowledges the DFG, the Rutgers University
Center for Materials Theory and NSF DMR0081075 for financial support
at early stages of this work; AD and AJM acknowledge NSF DMR0081075.

%%%%%%%%%%%%%%%%%%%%%%%%%%%%%%%%%%%%%%%%%%%%%%%%%%%%%%%%%%%%%%%%%%%%%%%
%%%%%%   REFERENCES   %%%%%%%%%%%%%%%%%%%%%%%%%%%%%%%%%%%%%%%%%%%%%%%%%
%%%%%%%%%%%%%%%%%%%%%%%%%%%%%%%%%%%%%%%%%%%%%%%%%%%%%%%%%%%%%%%%%%%%%%%

%\vspace*{-6mm}

\end{document}